\documentclass[prl,superscriptaddress,twocolumn,showpacs,amsmath]{revtex4}

\usepackage{bm}
\usepackage{graphicx}

\newcommand{\be}{\begin{equation}}
\newcommand{\ee}{\end{equation}}
\newcommand{\bea}{\begin{eqnarray}}
\newcommand{\eea}{\end{eqnarray}}
\newcommand{\beann}{\begin{eqnarray*}}
\newcommand{\eeann}{\end{eqnarray*}}
\newcommand{\besa}[1]{\begin{subequations}\label{#1}\begin{eqnarray}}
\newcommand{\eesa}{\end{eqnarray}\end{subequations}}
\newcommand{\bra}[1]{\langle #1 | \,}
\newcommand{\ket}[1]{\, | #1 \rangle}

\newcommand{\la}{\lambda}

\newcommand{\om}{\omega}
\newcommand{\Om}{\Omega}
\newcommand{\ga}{\gamma}
\newcommand{\Ga}{\Gamma}
\newcommand{\de}{\delta}
\newcommand{\De}{\Delta}
\newcommand{\ka}{\kappa}
\newcommand{\eps}{\epsilon}
\newcommand{\veps}{\varepsilon}
\newcommand{\br}{\mathbf{r}}
\newcommand{\bk}{\mathbf{k}}
\newcommand{\Ec}{\hat{\mathcal{E}}}
\newcommand{\lra}{\leftrightarrow}
\newcommand{\hsig}{\hat{\sigma}}

\begin{document}

\title{Quantum information processing with single photons and atomic ensembles 
\\in microwave coplanar waveguide resonators}

\author{David Petrosyan}
\affiliation{Institute of Electronic Structure \& Laser, 
FORTH, GR-71110 Heraklion, Crete, Greece}
\author{Michael Fleischhauer}
\affiliation{Fachbereich Physik, Universit\"at Kaiserslautern, 
D-67663 Kaiserslautern, Germany}

\date{\today}

\begin{abstract}
We show that pairs of atoms optically excited to the Rydberg states can 
strongly interact with each other via effective long-range dipole-dipole
or van der Waals interactions mediated by their non-resonant coupling to 
a common microwave field mode of a superconducting coplanar waveguide 
cavity. These cavity mediated interactions can be employed to generate 
single photons and to realize in a scalable configuration a universal phase 
gate between pairs of single photon pulses propagating or stored in 
atomic ensembles in the regime of electromagnetically induced
transparency. 
\end{abstract}


\maketitle


Ensembles of trapped atoms or molecules are promising systems for 
quantum information processing and communications \cite{QCcomp}.
They can serve as convenient and robust quantum memories for photons,
providing thereby an interface between static and flying qubits 
\cite{lukpet}, using e.g. stimulated Raman techniques, such as 
electromagnetically induced transparency (EIT) \cite{EITrev}. 
However, controlled interactions realizing universal quantum 
logic gates and entanglement in a deterministic and scalable way 
are difficult to achieve with photonic qubits propagating or stored 
in atomic ensembles.

A promising scheme for deterministic logic operations between stored
photonic qubits was proposed in \cite{LFCDJCZ}. The proposal exploits 
the so-called dipole blockade mechanism, wherein strong resonant 
dipole-dipole interaction (DDI) between Rydberg atoms suppresses 
multiple excitations within a certain interaction volume. First 
proof-of-principle experiments have impressively demonstrated the 
blockade effect for related van der Waals interactions (VdWIs) 
between Rydberg atoms \cite{vdwblk}. The achieved blockade radius 
of only a few $\mu$m is, however, not yet sufficient for implementing 
logic gates. Furthermore, the Rydberg blockade of \cite{LFCDJCZ} has 
two principle drawbacks. 
(i)~In free-space, the DDI scales with interatomic distance $r_{ij}$ 
as $r_{ij}^{-3}$. The blockade gap in the Rydberg excitation spectrum 
of an atomic cloud is determined by the smallest DDI between pairs 
of atoms at opposite ends of the cloud. Yet, for closely spaced atoms, 
the DDI can be very large, which may lead to level crossings with 
other Rydberg states opening detrimental loss channels.
(ii)~Complete excitation blockade in an atomic ensemble requires 
spherical symmetry of the resonant DDI, which severely restricts 
the choice of suitable Rydberg states. 

Here we put forward an alternative, scalable and efficient approach 
untainted by the above difficulties. We first show that superconducting 
coplanar waveguide (CPW) resonators \cite{BHWGS,MCGKJS-SPS}, operating in 
the microwave regime, can mediate long-range controlled interactions between
neutral atoms optically excited to the Rydberg states. By appropriate 
choice of the system parameters, effective resonant DDI or VdWI between 
pairs of atoms located near the CPW surface \cite{RDDLSZ} can be achieved. 
These interactions can then be employed to generate single photons and 
to realize a universal phase gate between pairs of single photon pulses 
propagating or stored in cold trapped atomic ensembles in the EIT regime.


\begin{figure}[b]
\includegraphics[width=0.48\textwidth]{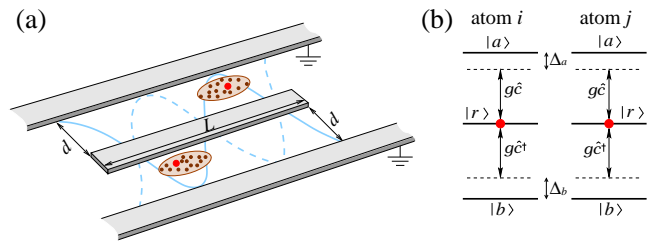}
\caption{(a)~CPW cavity with strip-line length $L$ and electrode
distance $d$. Ensembles of ground state atoms are trapped near 
the CPW surface within antinodes of the standing wave field. 
(b) Level scheme of a pair of excited Rydberg atoms $i$ and $j$ 
interacting with each other via non-resonant coupling to a common 
cavity mode.} 
\label{fig:slrvdwi}
\end{figure}

Consider a pair of atoms $i$ and $j$ optically excited to the Rydberg 
states $\ket{r}$. The atoms interact non-resonantly with a certain 
mode of CPW cavity with frequency $\om_c$ via transitions to adjacent 
Rydberg states $\ket{a}$ and  $\ket{b}$ lying, respectively, 
above and below $\ket{r}$ (Fig.~\ref{fig:slrvdwi}). All the 
other cavity modes are far detuned from the atomic transition frequencies 
$\om_{ar}$ and $\om_{rb}$ and do not play a role. In the frame rotating with
$\om_c$, the Hamiltonian is given by
\be
H =
\hbar \sum_{l= i,j} \big[ ( \De_{a} \hsig_{aa}^l - \De_{b} \hsig_{bb}^l ) 
+ ( g_{br}^l \hat{c}^{\dag} \hsig_{br}^l + 
g_{ar}^l \hat{c} \, \hsig_{ar}^l + \mathrm{H.c} ) \big] , \label{Ham}
\ee
where $\hsig_{\mu \nu}^l = \ket{\mu_l}\bra{\nu_l}$ is the 
transition operator of the $l$th atom, $\De_a = \om_{ar} - \om_c$ 
and $\De_b = \om_{rb} - \om_c$ are the corresponding detunings, 
$\hat{c}^{\dag}$ and $\hat{c}$ are the creation and annihilation
operators for the cavity field, and 
$g_{\mu \nu}^l = - (\wp_{\mu \nu} / \hbar) \veps_c u(\br_l)$ is the
atom-field coupling rate, which is determined by the dipole matrix element 
$\wp_{\mu \nu}$ of the atomic transition, the field per photon 
$\veps_c = \sqrt{\hbar \om_c/\eps_0 V_c}$ within the effective cavity
volume $V_c = 2 \pi d^2 L$, and the cavity mode function $u(\br_l)$ at 
atomic position $\br_l$ \cite{RDDLSZ,BHWGS}. 

Given an initial configuration $\ket{r_i}\ket{r_j}\ket{0_c}$, with 
both atoms in state $\ket{r}$ and zero cavity photons $\ket{0_c}$, 
and large detunings $|\De_{a,b}| \gg g_{ar}^l,g_{br}^l, \ka$, where 
$\ka$ is the cavity mode linewidth, we can use second order 
perturbation theory to eliminate the non-resonant states 
$\ket{r_{i,j}}\ket{b_{j,i}}\ket{1_c}$ with a single photon in the cavity.
We then obtain that each atom in state $\ket{r_l}$ experiences a 
cavity-induced level shift 
$s_r^l = \sqrt{|g_{br}^l|^2 + \De_b^2/4} - \De_b/2 \simeq |g_{br}^l|^2/\De_b$ 
(ac Stark shift) and small level broadening 
$\ga_r^l = \ka |g_{br}^l|^2/\De_b^2 \ll \ka$. In addition, state 
$\ket{r_i}\ket{r_j}$ couples to states $\ket{b_{i,j}}\ket{a_{j,i}}$
with rate $D_{ij} = g_{br}^{i} g_{ar}^{j}/\De_b = g_{ar}^{i} g_{br}^{j}/\De_b$
via virtual photon exchange between the atoms in the cavity 
\cite{BHWGS,RDDLSZ,BurImam,MCGKJS-SPS}. The corresponding 
interaction Hamiltonian reads 
\be
V_{ij}^{(2)}= \hbar D_{ij} \big( \hsig_{br}^{i} \hsig_{ar}^{j} +
\hsig_{ar}^{i} \hsig_{br}^{j}\big) +\mathrm{H.c} . \label{int2ham}
\ee
Note that states $\ket{r_i}\ket{r_j}\ket{0_c}$ and 
$\ket{b_{i,j}}\ket{a_{j,i}}\ket{0_c}$ are also coupled to state 
$\ket{b_i}\ket{b_j}\ket{2_c}$ with two photons in the cavity. 
But due to the large detunings $2 \De_b$, these transitions yield 
only small (fourth order) level shifts accounted for below. 
In second order in $g_{\mu \nu}^l$, the energy offsets 
of states $\ket{a_{i,j}}\ket{b_{j,i}}$, relative to $\ket{r_i}\ket{r_j}$, 
are $\hbar(\de \om + s_a^{i,j} - s_r^i - s_r^j)$, where 
$\de \om = \De_a -\De_b = \om_{ar} - \om_{rb}$ and 
$s_a^{i,j} = |g_{ar}^{i,j}|^2/\De_b$. If, by an appropriate choice of 
$\de \om$ (with $s_a^j = s_a^i$), the transitions 
$\ket{r_i}\ket{r_j} \lra \ket{b_{i,j}}\ket{a_{j,i}}$ 
are made resonant, the Hamiltonian (\ref{int2ham}) would describe an 
effective resonant DDI, or F\"oster process, between a pair of Rydberg 
atoms $i$ and $j$. Then the eigenstates of (\ref{int2ham}) form a
triplet of states $\ket{\psi_{ij}^0} 
= \frac{1}{\sqrt{2}} (\ket{b_i}\ket{a_j} - \ket{a_i}\ket{b_j} )$ and 
$\ket{\psi_{ij}^{\pm}}  =  \frac{1}{\sqrt{2}} \ket{r_i}\ket{r_j} 
\pm \frac{1}{2} (\ket{b_i}\ket{a_j} + \ket{a_i}\ket{b_j} )$
with the corresponding energies $0$ and $\pm \hbar \sqrt{2} D_{ij}$
relative to that of state $\ket{r_i}\ket{r_j}$. Note that unlike
the free space DDI of \cite{LFCDJCZ}, here the DDI has very long 
range as it is mediated by the cavity field extending over $L \sim 1\:$cm. 

We next consider the non-resonant case of $|\de \om| \gg D_{ij},s_{\mu}^l$. 
Starting from Hamiltonian (\ref{Ham}), we use fourth order perturbation
theory to eliminate states $\ket{r_{i,j}}\ket{b_{j,i}}\ket{1_c}$, 
$\ket{b_{i,j}}\ket{a_{j,i}}\ket{0_c}$ and $\ket{b_i}\ket{b_j}\ket{2_c}$,
connected to the initial state $\ket{r_i}\ket{r_j}\ket{0_c}$ via 
non-resonant single- and two-photon transitions. This yields 
\be
V_{ij}^{(4)}= \hbar \, \hsig_{rr}^{i} W_{ij} \hsig_{rr}^{j} , \label{int4ham}
\ee
which describes an effective cavity mediated VdWI between a pair of 
Rydberg atoms $i$ and $j$ with strength
$W_{ij} = 2 |g_{br}^i|^2 |g_{br}^j|^2/\De_b^3 - 
2 |g_{br}^i|^2 |g_{ar}^j|^2/(\de \om \De_b^2)$. 
The effect of this interaction is to shift the energy of two atoms 
simultaneously excited to state $\ket{r}$ by the amount $W_{ij}$.

Before proceeding, let us estimate the relevant parameters achievable 
in a realistic experiment. For atoms placed in the vicinity of CPW field 
antinodes, the coupling constants $g_{\mu \nu}^l$ are approximately 
the same, $g_{br}^l \approx g_{ar}^l \equiv g_r$ (see below). Setting 
$\De_{a,b} \simeq g_r f$ ($f \gg 1$) and $\de \om = s_r = g_r f^{-1}$,
the DDI coefficient is $D_{ij} \simeq D = g_r f^{-1}$. 
On the other hand, with $\De_{b} \simeq g_r f $ and $\de \om = - g_r $ 
[$\De_a \simeq g_r (f-1)$], the VdWI strength is 
$W_{ij} \simeq W = 4 g_r f^{-3}$. The total relaxation rate of a 
Rydberg state $\ket{\mu}$ ($\mu = r,a,b$) is $\Ga_{\mu} + \ga_{\mu}$, 
where $\Ga_{\mu}$ is a small intrinsic decay and 
$\ga_{\mu} \simeq \ka f^{-2}$ is the cavity-induced relaxation. 
To achieve coherent interactions, we require that 
$D,W \gg \Ga_{\mu} + \ga_{\mu}$, which leads to the condition 
$g_r >  \Ga_{r} f^3, \ka f$, tantamount to the strong coupling regime 
of cavity QED \cite{RBHrev}. 

For a CPW cavity with strip-line length $L \simeq 1\:$cm and 
electrode distance $d \simeq 15\:\mu$m [Fig.~\ref{fig:slrvdwi}(a)], 
the effective cavity volume is $V_c \simeq 1.4 \times 10^{-11}\:$m$^3$. 
The mode functions are 1D standing waves 
$u(z) = \cos (m \pi z/L)$ or $\sin (m \pi z/L)$ with $m$ 
being, respectively, an even or odd integer and $z \in [- L/2, L/2]$ 
\cite{BHWGS,MCGKJS-SPS}. Choosing e.g. $m = 5$, the mode wavelength 
is $\la_c = 2L/m \simeq 4\:$mm and there are $m+1$ field 
antinodes. With effective dielectric constant $\eps_r \sim 6$,
the mode frequency $\om_c = 2\pi c/\la_c \sqrt{\eps_r} \simeq 
2\pi \times 30\:$GHz. For properly selected Rydberg states 
$\ket{r}, \ket{a}, \ket{b}$, the transition frequencies 
$\om_{ar},\om_{rb}\sim \om_c$ can be adjusted with high precision using 
static electric and magnetic fields \cite{RydAtoms}. The dipole 
matrix element between neighboring Rydberg states with principal quantum 
number $n$ scales as $\wp \propto n^2 a_0 e$, which for $n \sim 50$ 
yields $g_r \sim 2\pi \times 10\:$MHz. In a cavity with quality 
factor $Q \simeq 10^6$, the photon decay rate is 
$\ka = \om_c /Q \simeq 200\:$KHz, while $\Ga_{\mu} \lesssim 1\:$KHz. 
Thus the strong-coupling regime with the above stringent condition can 
be realised for $f < \min \big[\sqrt[3]{g_r/\Ga_r},  g_r/\ka \big] \sim 40$. 

\begin{figure}[t]
\includegraphics[width=0.48\textwidth]{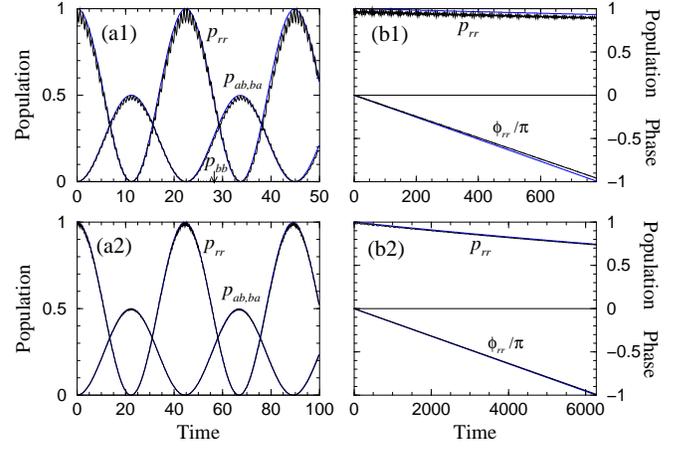}
\caption{Dynamics of two Rydberg atoms $i$ and $j$ 
coupled through a CPW cavity, initially in state 
$\ket{r_i}\ket{r_j}\ket{0_c}$.
(a1),(a2) correspond to effective DDI, and (b1),(b2) to 
effective VdWI. All parameters are given in the text, 
$f=10$ in (a1),(b1), and $f=20$ in (a2),(b2). 
$p_{\mu\nu}$ are populations of states $\ket{\mu_i}\ket{\nu_j}$, 
and $\phi_{rr}$ is the phase shift of $\ket{r_i}\ket{r_j}$. 
Thin (black) lines are numerical solutions of the master equation
for the full system using Hamiltonian (\ref{Ham}), and thick (blue) 
lines are solutions for the corresponding reduced model described 
by Eq.~(\ref{int2ham}) or (\ref{int4ham}). 
Time is measured in units of $g_r^{-1}$.} 
\label{fig:hameff}
\end{figure} 

Employing a master equation approach \cite{QCcomp} with the exact 
Hamiltonian (\ref{Ham}), we have performed numerical simulations 
of the dissipative dynamics of the full system with above 
parameters and $f=10,20$. Results for the cases realizing  
the effective DDI and VdWI are shown in Fig.~\ref{fig:hameff} 
and compared to simulations with the corresponding 
effective Hamiltonians (\ref{int2ham}) and (\ref{int4ham}). 
As expected, the agreement between the exact and effective models 
is good for $f=10$ and excellent for $f=20$. In the case of DDI, 
for both values of $f$, the decoherence and population losses are 
very small on the time scale of several oscillation periods 
$(\sqrt{2} D)^{-1}$ between states $\ket{r_i}\ket{r_j}$ and 
$\ket{b_{i,j}}\ket{a_{j,i}}$. In the case of VdWI, the population loss 
is appreciable on the much longer time scale of $W^{-1}$: At time 
$T_{\pi} = \pi/ W$, when state $\ket{r_i}\ket{r_j}$ acquires phase shift 
$\phi_{rr} = \pi$, its population $p_{rr} \simeq 0.92$ for $f =10$, 
and $p_{rr} \simeq 0.74$ for $f =20$. Thus, with realistic experimental 
parameters and $f = 10$ ($D \simeq 2\pi \times 1\:$MHz, 
$W \simeq 2\pi \times 40\:$KHz and $\ga \simeq 2\:$KHz), 
conditional phase shift of $\pi$ for a pair of Rydberg atoms can be 
achieved with fidelity $F > 90\%$. With CPW cavity improvements
and parameter optimization, the above fidelity may be further increased.

We envision several quantum information protocols utilizing 
Hamiltonians (\ref{int2ham}) and (\ref{int4ham}) in ensembles 
of alkali atoms in the ground state. Trapping cold atoms at a 
distance of $10$-$20\:\mu$m from the surface of a superconducting chip, 
incorporating the CPW cavity [Fig. \ref{fig:slrvdwi}(a)], is possible 
with presently available techniques \cite{atchip,mtrap,otrap}. 
A cigar shaped volume $V_a \sim d \times d \times \la_c /20$ would contain 
$N \sim 10^{6}$ atoms at density $\rho_0 \sim 2 \times 10^{13}\:$cm$^{-3}$. 
Each atomic ensemble should be positioned near the cavity field antinode, 
so that the mode function $|u(\br)| \approx 1$ is approximately constant 
throughout the atomic cloud. The corresponding coupling constants 
$g_{\mu\nu}$ can then be assumed the same for all the atoms in the CPW cavity. 


Employing light storage techniques based on EIT \cite{lukpet,EITrev}, 
the atomic ensembles in the setup of Fig.~\ref{fig:slrvdwi}(a) can 
serve as reversible quantum memories for single photon qubits.
Briefly, in a typical EIT setup, atoms in the ground state $\ket{g}$ 
resonantly interact with a weak (quantum) field $\Ec$ on the transition 
$\ket{g} \lra \ket{e}$, while a coherent driving field with Rabi frequency 
$\Om_d$ (and wavevector $\bk_d$) couples the excited state $\ket{e}$ to the 
long-lived (metastable) state $\ket{s}$. When the light pulse $\Ec$ 
(with wavevector $\bk$) enters the EIT medium, it is transformed into 
the so-called dark-state polariton 
$\Psi = \cos \theta \, \Ec - \sin \theta \, \sqrt{N} \hsig_{gs}$ 
\cite{fllk,EITrev}, which propagates in the medium with reduced group 
velocity $v_g = c \cos^2 \theta$, where $\tan \theta = g_{ge} \sqrt{N}/ \Om_d$ 
with $g_{ge} = \wp_{ge} \sqrt{\om / (2 \hbar \eps_0 V_a)}$. 
The slowing down of the pulse upon entering the medium leads to its 
spatial compression by a factor of $\cos^2 \theta \ll 1$ 
($0 < \theta \lesssim  \pi/2$). Once the pulse has been fully accommodated 
in the medium, by turning off $\Om_d$ ($\theta = \pi/2$), the photonic 
excitation is adiabatically mapped onto the collective 
long-lived atomic excitation represented by state 
$\ket{s^{(1)}} \equiv  1/\sqrt{N} \sum_{i=1}^N e^{i (\bk - \bk_d) \cdot \br_i} 
\ket{g_1,\ldots,s_i, \ldots,g_N}$ which involves a single Raman (spin) 
excitation, i.e., atom in the metastable state $\ket{s}$. At a later 
time, the photon can be retrieved on demand by turning $\Om_d$ on. 
Importantly, in order to accommodate the pulse in the medium with 
negligible losses, the optical depth of the atomic cloud should 
be large \cite{fllk,EITrev}. With a typical resonant absorption 
cross-section for the alkali atoms $\sigma_0 \simeq 10^{-10}\:$cm$^{2}$,
and the above cited density $\rho_0$ and medium length 
$L_a = \la_c/20 \sim 0.2\:$mm, we have large optical depth  
$2 \sigma_0 \rho_0 L_a \simeq 80$. 

\begin{figure}[t]
\includegraphics[width=0.48\textwidth]{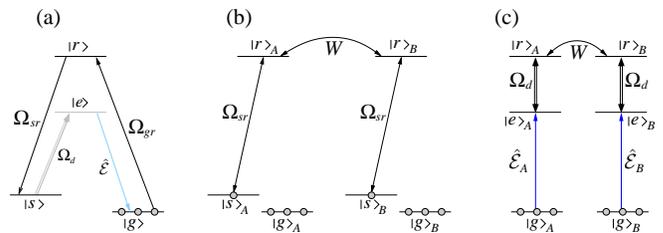}
\caption{(b)~Cavity-mediated DDI between atoms in Rydberg states 
$\ket{r}$ facilitates generation of single collective spin excitation,
via sequential application of $\Om_{gr}$ and $\Om_{sr}$ pulses. 
This excitation can then be converted into a single photon pulse $\Ec$
by applying $\Om_d$.
(b) Pair of atoms from ensembles $A$ and $B$ excited to states 
$\ket{r}$ interact via the cavity-mediated VdWI $W$ resulting in a 
conditional phase shift for the two ensemble qubits.
(b) Two single-photon fields $\Ec_A$ and $\Ec_B$ upon entering the 
corresponding atomic ensembles in the EIT regime are converted into 
two dark-state polaritons which interact with each other via VdWI 
and acquire a dispersive phase shift.} 
\label{fig:qlcmb}
\end{figure}


Using the cavity-mediated DDI (\ref{int2ham}), we can implement the 
dipole blockade \cite{LFCDJCZ} of multiple Rydberg excitations
in an atomic ensemble. This, in turn, can be used to prepare the 
collective state $\ket{s^{(1)}}$ and subsequently generate single 
photon pulses, as summarized below. Consider the level scheme of 
Fig.~\ref{fig:qlcmb}(a), where coherent laser fields with Rabi frequencies 
$\Om_{gr}$ and $\Om_{sr}$ and wavevectors $\bk_{gr}$ and $\bk_{sr}$ resonantly 
couple the lower atomic states $\ket{g}$ and $\ket{s}$ to the Rydberg state 
$\ket{r}$. Initially all atoms are in state $\ket{g}$, and only $\Om_{gr}$ 
is on. Then the laser field induces the transition from the 
ground state $\ket{g_1,g_2,\ldots,g_N} \equiv \ket{s^{(0)}}$ to the 
collective state $\ket{r^{(1)}} \equiv 1/\sqrt{N} \sum_i 
e^{i \bk_{gr} \cdot \br_i} \ket{g_1,\ldots,r_i, \ldots,g_N}$ representing
a symmetric single Rydberg excitation of the atomic ensemble. The
collective Rabi frequency for transition $\ket{s^{(0)}} \to \ket{r^{(1)}}$
is $\sqrt{N} \Om_{gr}$. Once an atom $i\,(\in \{1,\ldots,N \})$ is 
transferred to state $\ket{r}$, the excitation of a second atom 
$j\,(\neq i)$ is suppressed by the resonant DDI between the atoms, 
provided that $D_{ij} \simeq D > \Om_{gr}$. Indeed, out of the three 
eigenstates of (\ref{int2ham}), the unshifted eigenstate $\ket{\psi_{ij}^0}$ 
is not coupled to state $\ket{r_i} \ket{g_j}$ by $\Om_{gr}$, while  
transitions $\ket{r_i} \ket{g_j} \to \ket{\psi_{ij}^{\pm}}$ are shifted 
away from resonance by $\pm \sqrt{2} D_{ij}$ and therefore are inhibited.
Hence, a laser pulse of area $\sqrt{N} \Om_{gr} T_1 = \pi /2$ 
(an effective $\pi$ pulse) prepares the state $\ket{r^{(1)}}$. 
The probability of error due to populating the doubly-excited states 
$\ket{\psi_{ij}^{\pm}}$ is found by adding the probabilities of all possible 
double-excitations, $P_{\rm double} \sim N^{-1} \sum_{i,j} 2 |\Om_{gr}|^2/ (4 D_{ij}^2)
\approx N |\Om_{gr}|^2/(2 D^2)$. Additionally, Rydberg state relaxation 
during the pulse time $T_1 = \pi /(2 \sqrt{N} \Om_{gr})$ causes an
error $P_{\rm decay} \lesssim (\Ga_r + \ga_r) T_1$. The total error 
probability is minimized by choosing 
$\Om_{gr} = \sqrt[3]{\pi (\Ga_r + \ga_r) D^2/(2N^{3/2})}$. By subsequent 
application of the second stronger laser with pulse area $\Om_{sr} T_2 = \pi /2$ 
($\pi$ pulse), state $\ket{r^{(1)}}$ is quickly converted into 
$\ket{s^{(1)}} \equiv \frac{1}{\sqrt{N}} 
\sum_i e^{i (\bk_{gr} - \bk_{sr}) \cdot \br_j} \ket{g_1,\ldots,s_i, \ldots,g_N}$,
which is precisely the state we need for generating a single photon,
as described above and illustrated in Fig.~\ref{fig:qlcmb}(a). 
For the present parameters and choosing the optimal
$\Om_{gr} \simeq 2\pi \times 100\:$Hz, the preparation time of state 
$\ket{s^{(1)}}$ is $T \sim 2.7\:\mu$s with the fidelity  
$1-(P_{\rm double} + P_{\rm decay}) \gtrsim 0.98$.

Note that Rydberg atoms interact also via direct DDI 
$\bar{D}_{ij} \simeq \wp_{rb}\wp_{ra}/(4 \pi \eps_0 \hbar r_{ij}^3)$ \cite{LFCDJCZ},
which, for closely spaced atoms $i$ and $j$, is much larger than the 
cavity-mediated DDI $D_{ij}$. However, $\bar{D}_{ij}$ is a short-range
interaction, and already at interatomic distances $r_{ij} \gtrsim 20\:\mu$m 
it is smaller than $D_{ij}$, whose range is given by the CPW cavity size.

Since the cavity-mediated DDI is present between any pair of Rydberg atoms 
in the cavity, the above technique can be extended to create an entangled 
state of any two ensembles $A$ and $B$ within the cavity of 
Fig.~\ref{fig:slrvdwi}(a). Thus by applying $\Om_{gr}$ simultaneously 
to both ensembles, due to the dipole blockade, only one atom will be 
excited to state $\ket{r}$. The duration of the pulse should 
be chosen according to $\sqrt{2N} \Om_{gr} T_1= \pi/2$, since it now 
drives $2N$ atoms. Using the second laser pulse with area 
$\Om_{sr} T_2 = \pi /2$ to quickly bring the population 
of state $\ket{r}$ to the metastable state $\ket{s}$, an 
entangled state $(\ket{s^{(1)}}_A \ket{s^{(0)}}_B 
+ \ket{s^{(0)}}_A \ket{s^{(1)}}_B )/\sqrt{2} $ of atomic ensembles 
$A$ and $B$ sharing a single collective spin excitation, will be produced.


We now describe possible uses of the cavity-mediated VdWI (\ref{int4ham}). 
As detailed above, atomic ensembles can serve as reversible quantum 
memories for photonic qubits. Conversely, individual ensembles 
can represent qubits storing any state of the form 
$\ket{\psi} = \alpha \ket{0} + \beta \ket{1}$ in the corresponding 
superposition $\ket{\psi} = \alpha \ket{s^{(0)}} + \beta \ket{s^{(1)}}$
of collective states. Consider two such ensemble qubits $A$ and $B$
in the cavity of Fig. \ref{fig:slrvdwi}(a). A resonant $\pi$-pulse 
applied to the transition $\ket{s} \to \ket{r}$ in both atomic 
ensembles, $\Om_{sr} T_1 = \pi/2$, will then convert state 
$\ket{s^{(1)}})_{A,B}$ of each ensemble to the state $\ket{r^{(1)}}_{A,B}$ 
with single Rydberg excitation [see Fig. \ref{fig:qlcmb}(b)]. 
Since any two atoms in state $\ket{r}$ interact via VdWI with
strength $W$, during time $T_{\pi} = \pi /W$ they will accumulate a 
phase shift $\pi$. Only if both ensembles were initially in state 
$\ket{s^{(1)}}$, the above phase shift occurs, since otherwise 
there will be only zero or one atom in state $\ket{r}$ and the VdWI 
(\ref{int4ham}) will not take place. A second $\pi$-pulse 
$\Om_{sr} T_2 = \pi/2$ can then convert state $\ket{r^{(1)}}_{A,B}$ 
of each ensemble back to the original state $\ket{s^{(1)}}_{A,B}$. 
Thus the universal \textsc{cphase} gate 
$\ket{s^{(x)}}_A \ket{s^{(y)}}_B \to (-1)^{xy}\ket{s^{(x)}}_A \ket{s^{(y)}}_B$
($x,y \in [0,1]$) \cite{QCcomp} would be implemented between  
$A$ and $B$.

The VdWI between Rydberg atoms can also be employed to implement the 
\textsc{cphase} gate directly between two photonic qubits. A possible 
setup is shown of Fig.~\ref{fig:qlcmb}(c), where single photon 
fields $\Ec_A$ and $\Ec_B$ propagate in the atomic ensembles $A$ 
and $B$ under the conditions of EIT in the ladder configuration: 
the quantum fields act on transition $\ket{g} \to \ket{e}$, 
while transition $\ket{e} \to \ket{r}$ is driven by resonant 
classical field with Rabi frequency $\Om_d$. Upon entering the medium, 
each quantum field is converted into the dark-state 
polariton $\Psi_{A,B} = \cos \theta \, \Ec_{A,B} 
- \sin \theta \, \sqrt{N} \hsig_{gr}^{A,B}$ whereby part of the 
photonic excitation is temporally transferred to the atomic excitation. 
For $\theta \lesssim \pi/2$, the polariton propagating with group velocity
$v_g = c \cos^2 \theta \ll c$ is mainly an atomic excitation.
Thus, each field $\Ec_{A,B}$ containing a single photon creates in the 
corresponding ensemble an atom in state $\ket{r}$. These atoms interact via 
the cavity mediated VdWI (\ref{int4ham}) resulting in the cross-phase 
modulation between the two quantum fields. If both fields enter the 
corresponding ensembles simultaneously, the interaction time is 
$T_{\mathrm{int}} = L_a/v_g$. Calculations similar to those in \cite{DDEIT} 
show that the non-linear phase shift for a pair of single photons is 
given by $\phi = \sin^4 \theta \, W L_a/v_g$. We can express the group 
velocity as $v_g = |\Om_d|^2/\sigma_0 \rho_0 \ga_{ge}$, where $\ga_{ge}$ 
is the coherence relaxation rate on the transition $\ket{g} \to \ket{e}$, 
with typical value for alkali atoms $\ga_{ge} \simeq 15\:$MHz. With the 
other parameters given above and choosing $\Om_d = 2\pi \times 1.1\:$MHz
($\sin^2 \theta \simeq 1$), we obtain $\phi = \pi$. Thus the two-photon 
input state $\ket{1}_A \ket{1}_B$ acquires a conditional phase shift of
$\pi$, and more generally, the \textsc{cphase} gate 
$\ket{x}_A \ket{y}_B \to (-1)^{xy}\ket{x}_A \ket{y}_B$ ($x,y \in [0,1]$)
between two photonic qubits is realized.


To summarize, ensembles of cold atoms trapped in the vicinity of 
a microwave CPW cavity can strongly interact with each other via 
cavity mediated virtual photon exchange between optically exited 
atomic Rydberg states. This system can serve as efficient and 
scalable platform to realize various quantum information 
processing protocols with ensemble qubits and single photons.   

\begin{acknowledgments}
This work was supported by the 
EC Marie-Curie Research Training Network EMALI.
\end{acknowledgments}

\end{document}